\documentclass[11pt,a4paper]{article}

\usepackage[utf8]{inputenc}
\usepackage[T1]{fontenc}
\usepackage[english]{babel}
\usepackage{lmodern}
\usepackage{microtype}
\usepackage{geometry}
\usepackage{booktabs}
\usepackage{tabularx}
\usepackage{array}
\usepackage{graphicx}
\graphicspath{{figures/}}
\usepackage{hyperref}
\usepackage{enumitem}

\hypersetup{
  unicode=true,
  colorlinks=true,
  linkcolor=blue,
  citecolor=blue,
  urlcolor=blue,
  pdftitle={From Prompt to Process: a Process Taxonomy and Comparative Assessment of Frameworks Supporting AI Software Development Agents},
  pdfauthor={Sanderson Oliveira de Macedo}
}

\newcolumntype{Y}{>{\raggedright\arraybackslash}X}
\newcommand{\doi}[1]{\href{https://doi.org/#1}{doi:\ #1}}

\title{From Prompt to Process: a Process Taxonomy and Comparative Assessment of Frameworks Supporting AI Software Development Agents}
\author{Sanderson Oliveira de Macedo\\
\textit{Federal Institute of Goias}\\
\texttt{sanderson.macedo@ifg.edu.br}}
\date{\today}

\begin{document}

\maketitle

\begin{abstract}
AI tools for programming are no longer just autocomplete or chat assistants: they organize themselves as development frameworks, with process, roles, artifacts and verification. Recent surveys map agents and LLMs for software engineering, but a study centered on the operational frameworks that turn these capabilities into process is missing. We ran a directed search of primary sources, with a functional inclusion criterion and traction measurement, and selected six frameworks: GitHub Spec Kit, OpenSpec, BMAD Method, Get Shit Done (GSD), Spec Kitty and Reversa. Each attacks AI development through a different path: spec-driven development in full and lightweight variants, agent-driven agile planning, context engineering over the agent, worktree isolation and review, and recovery of operational specifications from legacy systems. Our central contribution is a six-dimension process taxonomy: specification, context, roles, execution, validation and portability, with a scoring rubric that turns it into a replicable instrument. We apply it to the six frameworks and an out-of-sample case, Spec-Flow. Two results stand out. Among frameworks that already adopt some process there is convergence: the isolated prompt loses centrality, and persistent artifacts, work contracts, traceability and human review become mechanisms that reduce ambiguity and coordinate agents. And no framework strongly covers all six dimensions, exposing a structural trade-off between process depth and portability across agents. We also found recurring risks: drift between specification and code, excessive trust in generated artifacts, fragility of community extensions, platform dependence and a lack of benchmarks for the complete process. We close with a research agenda for empirical evaluation, focused on intermediate-quality metrics, context governance, installation security and reproducibility.
\end{abstract}


\section{Introduction}

Software development advances through leaps in abstraction and automation. High-level languages, integrated development environments, version control, automated testing and continuous integration have, step by step, moved part of the manual effort to the tool. Language models continue this trajectory. AI started as autocomplete and pointwise code generation, supporting the programmer without changing how the work is organized \cite{hou2024llmse}.

Then came command-line development agents: Claude Code, OpenAI Codex CLI and Gemini CLI. They run an agentic loop. They receive an objective, plan, edit files, run commands and tests, read the result and iterate until done \cite{liu2024agentsurvey,jin2024llmagents}. The unit of interaction is no longer the isolated prompt; it becomes an autonomous session over the repository. The agent alone does not guarantee process. Without specification, traceability, defined roles and validation, the session reproduces, at a larger scale, the problems of pointwise interaction: lost context, unrecorded decisions, difficult review \cite{wang2024agentsse}.

Frameworks supporting AI software development attack this gap. We call a framework a structured set of artifacts, commands, roles, templates, workflows or policies that runs over the agent to organize whoever uses it. Using the agent directly is one thing; a framework is another: it defines how to collect context, how to record a decision, what role the agent takes at each step, when a human reviews and what evidence remains. The recent examples vary widely. The BMAD Method organizes planning, architecture and implementation into phases and specialized agents \cite{bmad2026docs,bmad2026github}. GitHub Spec Kit operationalizes Spec-Driven Development and treats specifications as a central artifact that feeds plans, tasks and implementation \cite{github2026speckitdocs,github2026speckitrepo}. Reversa starts from the other end: it uses reverse documentation engineering to turn legacy software into operational specifications that the agent consumes \cite{macedo2026reversa}.

Despite the popularity, there is no consensus on how to compare these frameworks. Some are methodologies, others toolkits, others community extensions. Some start from product and requirements; others, from legacy code; others, from commands and workflows over the agent. Recent surveys organized the field of software agents by tasks, benchmarks and internal components \cite{liu2024agentsurvey,wang2024agentsse,jin2024llmagents}; practical guides compare Spec-Driven Development tools by market criteria \cite{daniel2026sddpatterns,openspec2026comparison}. But comparing by number of agents, number of commands or repository stars is not enough: none of this captures the function each mechanism plays in the engineering process. This article fills the gap with four contributions: a six-dimension process taxonomy that makes portability across agents an explicit criterion, absent from product comparisons; its application to a set of frameworks of greatest traction that confronts, in the same analysis, greenfield specification-driven development and the recovery of specifications from legacy systems through reverse engineering; a map of recurring risks; and an empirical research agenda. We detail each one at the end of the section.

Our hypothesis is that the central value of these frameworks lies less in automating the writing of code and more in how they structure the cognitive and operational work around the agent. The question stops being ``which model writes better code?'' and becomes ``which process lets humans and agents keep context, traceability and control over changes?''. This brings the frameworks back to classic concerns of software engineering: specification, traceability, architecture, review, validation and maintenance \cite{tian2021traceability,hou2024llmse}.

The research questions that guide this study are:

\begin{itemize}[leftmargin=*]
  \item \textbf{RQ1:} Which architectural dimensions distinguish frameworks supporting AI software development?
  \item \textbf{RQ2:} Which frameworks that help users of development agents to develop software with AI can be identified by a directed search of primary sources and recent grey literature?
  \item \textbf{RQ3:} Which risks and gaps remain when these frameworks are used in real projects?
  \item \textbf{RQ4:} Which research agenda can support empirical evaluation, comparison and governance of these frameworks?
\end{itemize}

There are four contributions. The first is a six-dimension process taxonomy to analyze frameworks supporting AI software development. One dimension is portability across agents, which existing product comparisons do not make explicit. A scoring rubric accompanies the taxonomy and turns it into an instrument replicable from explicit criteria, not just a descriptive vocabulary. The second contribution applies this instrument to a set of frameworks of greatest traction. The set covers two extremes that rarely appear in the same analysis: greenfield specification-driven development (Spec Kit, OpenSpec, BMAD, GSD and Spec Kitty) and the recovery of operational specifications from legacy systems through reverse documentation engineering (Reversa). We also include an out-of-sample case, Spec-Flow, to test the generality of the instrument. The central result is simple: no framework strongly covers all six dimensions. There is a structural trade-off between process depth and portability. The third contribution discusses recurring risks of specification, autonomy, extensibility and evaluation. The fourth draws a research agenda for the field to move from demonstrations and usage reports to empirical evaluations of the complete process.

\section{Method}\label{sec:metodo}

We conduct a comparative study of frameworks supporting AI software development, grounded in a directed and qualitative search of primary sources rather than in an exhaustive systematic review. The study combines formal literature and grey literature under explicit criteria. We follow the guidance of Garousi et al. \cite{garousi2019mlr} on grey literature in software engineering and the criteria-transparency principles of Kitchenham et al. \cite{kitchenham2009slr}, but we do not claim the coverage nor the complete protocol of a systematic review. The formal literature (articles and preprints) grounds the state of the art and the discussion of risks. The grey literature (official documentation, repositories and public comparisons) is the primary source about the object-frameworks, which in general still lack independent academic evaluation. Our object is not the entire field of agents in software engineering, but a delimited set of support frameworks. That is why the selection follows a directed and purposeful search of primary sources, described below, rather than a sweep by strings in bibliographic databases. The selection is non-exhaustive: it prioritizes depth of characterization over coverage of the ecosystem.

\noindent\textbf{Functional inclusion criterion.} We fixed the unit of analysis before the selection, to avoid an arbitrary list of products. An artifact enters as an object-framework when it satisfies, at the same time, the four conditions below:

\begin{enumerate}[leftmargin=*]
  \item \textbf{Process-support function.} The artifact supports the AI software development cycle through specification, planning, context organization, roles, workflows or validation, and not only the pointwise generation of code.
  \item \textbf{Centered on the user of an agent.} The artifact is used by a developer who already operates a development agent (for example, Claude Code, OpenAI Codex CLI or Gemini CLI); the agent is the environment, and the framework is the support tool that runs over it.
  \item \textbf{Not the agent, IDE or closed platform itself.} Excluded are the agents or runtimes themselves and the proprietary IDEs or platforms that embed the agent, since they are not reusable support layers over an external agent.
  \item \textbf{Not an agent-building SDK.} Excluded are general-purpose libraries and SDKs for building agent systems, since they serve agent engineering and not the direct support of software development by those who use an agent.
\end{enumerate}

By the criterion, we excluded by category: the agents or runtimes (Claude Code, Codex, Gemini CLI), which are the environment; the closed IDEs and platforms (among them Google Antigravity and Kiro); and the agent-building SDKs (CrewAI, Agno, Google ADK, LangChain, LangGraph and AutoGen).

\noindent\textbf{Sources of the directed search.} The directed search went through primary and secondary sources of the current ecosystem of support frameworks: the documentation and the official repository of each candidate; community-curated lists (including \textit{awesome} collections on specification-driven development and on extensions for development agents); and public comparisons of Spec-Driven Development tools \cite{daniel2026sddpatterns,openspec2026comparison,bcms2026sddguide,augment2026sddtools}. We confronted each candidate raised against the functional inclusion criterion. Reversa, besides being a formally published object of study \cite{macedo2026reversa} and already part of this study's formal-literature corpus, meets the traction threshold described below; because it is authored by the author of this study, the corresponding conflict of interest is declared at the end of the section.

\noindent\textbf{Traction measurement.} A second filter keeps the selection from reflecting only editorial visibility: traction, which requires adoption above an explicit floor and recent activity. We included the candidate with at least one thousand GitHub stars and at least one push in the last six months; whoever fell below the star threshold or had no activity in the period was dropped. We measured traction via the GitHub API between May 26 and 28, 2026, recording the number of stars and the date of the last activity of each repository. The snapshot is in \texttt{notes/traction\_snapshot.csv}, for reproducibility. Table~\ref{tab:tracao} summarizes the decision for each candidate.

\begin{table}[ht]
\centering
\caption{Traction triage of the candidates (GitHub, May 2026).}
\label{tab:tracao}
\begin{tabularx}{\textwidth}{p{4.6cm}r p{2.4cm} Y}
\toprule
\textbf{Candidate (repository)} & \textbf{Stars} & \textbf{Activity} & \textbf{Decision} \\
\midrule
GitHub Spec Kit (\texttt{github/spec-kit}) & 106786 & active & included \\
Get Shit Done, GSD (\texttt{gsd-build/get-shit-done}) & 63754 & active & included \\
OpenSpec (\texttt{Fission-AI/OpenSpec}) & 51404 & active & included \\
BMAD Method (\texttt{bmad-code-org/BMAD-METHOD}) & 48209 & active & included \\
Spec Kitty (\texttt{Priivacy-ai/spec-kitty}) & 1273 & active & included (at the threshold) \\
Reversa (article + framework) & 1100 & active & included (at the threshold) \\
\midrule
claude-code-spec-workflow & 3748 & no push since 2025-09 & excluded for inactivity \\
Spec-Flow (\texttt{marcusgoll/Spec-Flow}) & 85 & active & excluded for low traction \\
Tessl (tile SDD) & 38 & commercial & excluded; kept as an ecosystem reference \\
\bottomrule
\end{tabularx}
\end{table}

\begin{figure}[h!]
\centering
\includegraphics[width=1.0\textwidth]{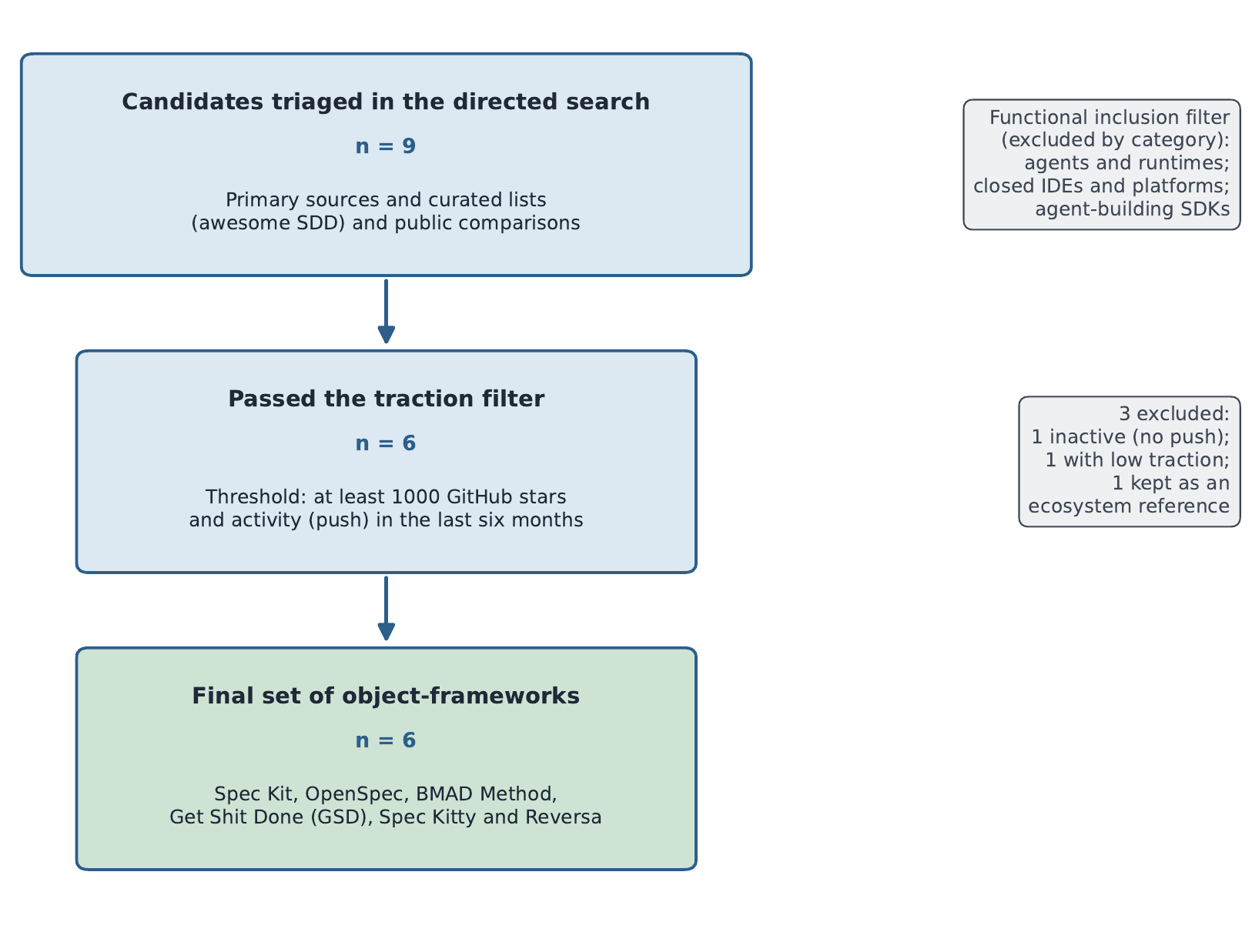}
\caption{Funnel of the directed search: candidates raised in primary and secondary sources, functional inclusion filter, traction filter and final set of six object-frameworks.}
\label{fig:funil}
\end{figure}

The final set has six object-frameworks: GitHub Spec Kit, OpenSpec, BMAD Method, Get Shit Done (GSD), Spec Kitty and Reversa. The candidates below the threshold or inactive were recorded as excluded, with the respective justification. Figure~\ref{fig:funil} summarizes the funnel of the directed sweep, from the survey of candidates to the final set.

\noindent\textbf{Extraction and synthesis.} We characterized each framework of the final set from its official documentation and repository, according to the six dimensions of the taxonomy proposed in Section~\ref{sec:taxonomia} (specification, context, roles, execution, validation and portability). The synthesis is comparative and qualitative. Section~\ref{sec:discussao} confronts the six frameworks dimension by dimension, points out recurring patterns and maps risks. We treat the productivity claims from the primary sources qualitatively; when they lack independent evaluation, they become a limitation, never a number of our own.

\noindent\textbf{Source confidence.} Most object-frameworks have not gone through independent academic evaluation. So we classify the evidence base of each one, to make the bias risk of each characterization explicit. Table~\ref{tab:confianca} records, per framework, the type of primary source, whether independent academic evaluation exists, the maturity (traction) and a confidence level. That level is the authors' judgement, not a measurement: it only weights each source in the synthesis. Lower-confidence sources, such as product documentation without external evaluation or low-traction frameworks, enter as evidence to confront with the formal literature, not as conclusive proof.

\begin{table}[h!]
\centering
\caption{Confidence of the evidence base per object-framework. The confidence level is the authors' judgement from the sources, not a validated measurement.}
\label{tab:confianca}
\begin{tabularx}{\textwidth}{p{2.7cm} p{3.6cm} p{2.5cm} p{1.7cm} Y}
\toprule
\textbf{Framework} & \textbf{Primary source} & \textbf{Independent academic evaluation} & \textbf{Maturity} & \textbf{Confidence} \\
\midrule
GitHub Spec Kit & Official documentation and repository & No & High & Medium \\
OpenSpec & Official documentation and repository & No & High & Medium \\
BMAD Method & Official documentation and repository & No & High & Medium \\
Get Shit Done (GSD) & Official documentation and repository & No & High & Medium \\
Spec Kitty & Official documentation and repository & No & Low & Low \\
Reversa & Article (preprint) and framework & Partial (preprint, with declared conflict of interest) & Medium & Medium \\
\bottomrule
\end{tabularx}
\end{table}

\noindent\textbf{Limitations of the method.} The directed search prioritizes depth of characterization over exhaustiveness of the ecosystem; low-traction or very recent frameworks may have been left out. Measuring traction by stars and activity is an imperfect indicator of adoption and of quality, and unstable over time: the values in Table~\ref{tab:tracao} are a May 2026 snapshot and tend to change. Part of the primary sources is grey literature with possible promotional bias, which we mitigate by confronting the claims with the formal literature and by reporting uncertainties explicitly. The dimensional scoring in Section~\ref{sec:discussao} was done by a single rater, without a second independent coder; we do not report inter-rater reliability, which is a threat to the objectivity of the rubric and remains future work. We also declare a conflict of interest: the Reversa framework \cite{macedo2026reversa} is authored by the author of this study. It meets the traction threshold like the other included frameworks, but, to mitigate the conflict, it receives the same critical analysis and the same record of risks, without preferential treatment. The dimensional scores reported in Section~\ref{sec:discussao} express the authors' judgement from official documentation: they serve as a comparative instrument replicable from explicit criteria, not as a third-party-validated empirical measurement.

\section{Related Work}

The related literature is organized into three fronts: broad reviews on AI and agents in software engineering, architectural discussions about the agentic development life cycle, and practical guides on specification-driven tools. This section synthesizes the state of the art on these fronts, details the gaps and positions our study.

\noindent\textbf{What Exists in the Literature.} Agents and Large Language Models (LLMs) in engineering tasks have already produced several reviews. Hou et al. \cite{hou2024llmse} map, in a systematic review, the use of LLMs by life-cycle task. Liu et al. \cite{liu2024agentsurvey} and Wang et al. \cite{wang2024agentsse} propose taxonomies by the cognitive components of the agent (perception, memory and action) and reasoning capability. Jin et al. \cite{jin2024llmagents} examine the challenges of agents in the design, testing and maintenance phases, and He et al. \cite{JundaHe2025} review the internal architecture and multi-agent cooperation, drawing a roadmap to consolidate agentic systems.

In a process line, Bhati \cite{bhati2026agenticsdlc} formalizes the transition to the agentic life cycle (A-SDLC) and documents productivity gains under human delegation. On the infrastructure and governance side, Sengupta et al. \cite{sengupta2026harnesses} propose meta-engineering harnesses with continuous verification and agentic contracts for sustainable large-scale production.

In the field of Spec-Driven Development (SDD), Piskala \cite{piskala2026sdd} treats specifications as living development contracts in the era of AI assistants. At a more technical level, Taghavi and Bhavani \cite{taghavi2026speckitagents} show that probing and validation hooks coupled to Spec Kit strengthen agentic flows and reduce context-blindness and hallucinated APIs in corporate repositories.

The commercial grey literature completes the picture: guides by Daniel \cite{daniel2026sddpatterns}, OpenSpec \cite{openspec2026comparison}, BCMS \cite{bcms2026sddguide} and Augment Code \cite{augment2026sddtools} compare real tools by ease of installation, platform coupling and setup complexity.

\noindent\textbf{Comparison with Existing Reviews and Guides.} Confronted directly, these works delimit the space of this study. Liu et al. \cite{liu2024agentsurvey} and He et al. \cite{JundaHe2025} categorize agents by cognitive components and by multi-agent cooperation; we adopt an engineering-process lens and compare operational frameworks by specification, context, roles, execution, validation and portability. Bhati \cite{bhati2026agenticsdlc} and Sengupta et al. \cite{sengupta2026harnesses} deal, respectively, with the macro transition to the agentic life cycle and with continuous-verification harnesses; we keep the unit of analysis on the support framework that runs over the agent, not on the agent nor on the organization. The closest confrontation is with the Spec-Driven Development comparisons. Daniel \cite{daniel2026sddpatterns}, the BCMS guide \cite{bcms2026sddguide}, the Augment selection \cite{augment2026sddtools} and the comparison from the OpenSpec portal \cite{openspec2026comparison} already cover Spec Kit, BMAD and OpenSpec, but as product content: without an explicit taxonomy, without an auditable selection criterion and, in the OpenSpec case, from one of the very tools being compared. Against these guides, this article differs in three verifiable points. The first is a declared functional inclusion criterion and a declared traction filter. The second is a portability dimension across agents that product comparisons do not make explicit. The third is the inclusion of reverse documentation engineering (Reversa \cite{macedo2026reversa}) as a complementary path, from the legacy to the specification, that greenfield SDD guides do not cover. Piskala \cite{piskala2026sdd} and Taghavi and Bhavani \cite{taghavi2026speckitagents}, in turn, deepen, respectively, the conceptual basis of SDD and grounding hooks over Spec Kit; we situate these pointwise contributions within the broader comparative ecosystem of the six frameworks.

Table~\ref{tab:posicionamento} summarizes this positioning. It confronts the anchor reviews and the product comparisons with this study on four verifiable criteria. No prior work meets all four. This study is the only one that compares support frameworks by process while also covering portability across agents, legacy recovery and an auditable selection criterion, over six frameworks.

\begin{table}[h!]
\centering
\caption{Positioning of this study relative to the anchor reviews and the product comparisons. A cell is marked when the work meets the criterion.}
\label{tab:posicionamento}
\begin{tabularx}{\textwidth}{Y p{2.3cm} p{1.9cm} p{1.7cm} p{1.9cm}}
\toprule
\textbf{Work} & \textbf{Compares support frameworks by process} & \textbf{Portability across agents} & \textbf{Legacy recovery} & \textbf{Auditable selection criterion} \\
\midrule
Hou et al. \cite{hou2024llmse} & No & No & No & Yes \\
Liu, Wang \cite{liu2024agentsurvey,wang2024agentsse} & No & No & No & Yes \\
Jin, He \cite{jin2024llmagents,JundaHe2025} & No & No & No & Yes \\
Bhati \cite{bhati2026agenticsdlc} & No & No & No & Partial \\
Product comparisons \cite{daniel2026sddpatterns,bcms2026sddguide,augment2026sddtools,openspec2026comparison} & Partial & No & No & No \\
This study & Yes (six) & Yes & Yes & Yes \\
\bottomrule
\end{tabularx}
\end{table}

\noindent\textbf{Research Gaps.} Despite the rapid advance of the literature, two main gaps remain:

\begin{enumerate}
    \item \textbf{Abstraction versus Operational Process}: Current academic reviews prioritize the abstract analysis of agent architectures or the classification of pointwise techniques by isolated life-cycle tasks. A Software Engineering lens is missing to characterize operational frameworks that organize the development cycle at the repository and process level.
    \item \textbf{Greenfield isolated from legacy recovery and portability}: Current comparisons, academic and grey, almost always deal with greenfield specification-driven development, that is, creating specifications for new software. They rarely discuss two aspects that are decisive in real projects. The first is recovering specifications from legacy systems, the reverse path, from code to specification, proposed by frameworks such as Reversa \cite{macedo2026reversa}. The second is the portability of a framework across agents, which product comparisons rarely treat as an explicit criterion. No review integrates these two aspects, together with the recent community frameworks of context engineering and flow (GSD and Spec Kitty), into a single process taxonomy; and the grey-literature comparisons, although useful, tend to be partial or to carry promotional product bias.
\end{enumerate}

\noindent\textbf{Proposal of Our Project.} To fill these gaps, we propose a comparative study that characterizes the practical transition from the isolated prompt to a traceable development process. It is structured into three central contributions, plus a map of recurring risks:

\begin{itemize}
    \item \textbf{Architectural Process Taxonomy}: We propose a unified comparative framework based on six operational process dimensions (specification, context, roles, execution, validation and portability), treating portability across agents as a first-class comparative criterion, and not as an installation detail.
    \item \textbf{Conceptual Comparative Matrix}: We analyze and confront in an impartial way six open, academic and community frameworks (GitHub Spec Kit, OpenSpec, BMAD Method, GSD, Spec Kitty and Reversa) that span from greenfield specification-driven development to reverse documentation engineering of legacy systems, delimiting their strengths, trade-offs and dependencies.
    \item \textbf{Research Agenda for Empirical Evaluation}: We outline a research agenda focused on process benchmarks, supply chain security of agentic templates, mitigation of specification drift and context governance.
\end{itemize}

\section{Proposed Taxonomy}\label{sec:taxonomia}

Our taxonomy organizes AI development frameworks into six dimensions: specification, context, roles, execution, validation and portability. We chose these six because they appear repeatedly, even if under different names, in the six analyzed frameworks (Section~\ref{sec:frameworks}).

\noindent\textbf{Specification.} Specification is how the framework turns human intention into artifacts that guide the agent. At the weak extreme, the specification is just a prompt. At the strong extreme, it is a versioned set of requirements, acceptance criteria, plans, tasks, architecture and policies. Spec Kit makes this explicit: it organizes the cycle into commands such as constitution, specification, plan, tasks and implementation \cite{github2026speckitdocs,github2026speckitrepo}. BMAD also bets on progressive artifacts, such as product brief, PRD, architecture, epics, stories and code review \cite{bmad2026docs}. Reversa changes the origin of the specification. Instead of starting from a new product, it recovers operational specifications from existing systems \cite{macedo2026reversa}.

\noindent\textbf{Context.} Context is how the framework decides what the agent should know before acting: documents, code, history, architectural decisions, local rules and collected evidence. The recent literature on Spec Kit Agents shows that agents can go ``blind to context'' in large repositories, and that probing and validation hooks reduce architectural violations and API hallucinations \cite{taghavi2026speckitagents}. Each framework solves this in its own way. BMAD builds context progressively across phases \cite{bmad2026docs}; Reversa extracts it from the legacy and labels it by confidence \cite{macedo2026reversa}; GSD treats context assembly as an explicit engineering task, deciding what the agent reads and in what order before acting \cite{gsd2026github}.

\noindent\textbf{Roles.} Roles define specialization, authority and behavioral expectation. Multi-agent frameworks usually divide the work among personas: analyst, product manager, architect, developer, QA, reviewer. This division reduces prompt ambiguity and creates output expectations. BMAD is the most explicit case in the set, with default agents for analysis, product, architecture, development, UX and technical writing \cite{bmad2026docs}. Spec Kitty distributes the work into packages isolated by worktrees, with review and acceptance before the merge, and so inserts execution and control roles into the flow \cite{speckitty2026github}. The idea is not new: ChatDev and other multi-agent works already explored role decomposition before these frameworks \cite{qian2024chatdev}.

\noindent\textbf{Execution.} Execution says whether the framework only guides or also triggers tools, edits code, runs tests and touches interfaces. In modern agentic tools, this dimension is central: the agent does not recommend, it changes the environment. SWE-agent showed the weight of agent-computer interfaces in real software engineering tasks \cite{yang2024sweagent}; SWE-bench consolidated a benchmark on real GitHub issues \cite{jimenez2024swebench}. Among the support frameworks, Spec Kitty goes further: it isolates execution in git worktrees and lets agents implement work packages in a controlled way before review and merge \cite{speckitty2026github}.

\noindent\textbf{Validation.} Validation gathers the mechanisms that check whether what was produced is correct, complete and aligned with the context: checklists, automated tests, human review, evidence artifacts, screenshots, browser recordings, findings reports and readiness gates. Spec Kit brings analysis and checklist commands; BMAD, PRD review, readiness checks and code review; Reversa makes confidence and gaps explicit; Spec Kitty places review and acceptance before the merge \cite{github2026speckitrepo,bmad2026docs,macedo2026reversa,speckitty2026github}. This dimension is critical because the agent can produce a plausible result without real adherence to the system. Execution and validation can co-activate over the same mechanism: Spec Kitty's worktrees isolate the implementation (execution), while requiring review and acceptance before the merge is validation. We treat the six dimensions as complementary reading lenses, not as disjoint partitions; therefore the same feature may score in more than one dimension.

\noindent\textbf{Portability.} Portability says whether the framework depends on a specific vendor, IDE, model or format. Spec Kit declares compatibility with several agents and integrations, which makes it a relatively portable layer over code assistants \cite{github2026speckitdocs}. OpenSpec announces support for dozens of code assistants and is even more portable because of its low overhead \cite{openspec2026github}. BMAD also aims at installation in different AI-assisted tools \cite{bmad2026github}, and Reversa emphasizes support for multiple coding engines \cite{macedo2026reversa}. On the other side, GSD is strongly coupled to a specific agent, which reduces its portability \cite{gsd2026github}; and Spec Kitty, although multi-agent, depends on its own repository conventions and on git worktrees \cite{speckitty2026github}.

\begin{table}[ht]
\centering
\caption{Dimensions of the proposed taxonomy.}\label{tab:dimensoes}
\begin{tabularx}{\textwidth}{p{2.8cm}Y Y}
\toprule
\textbf{Dimension} & \textbf{Guiding question} & \textbf{Observable indicators} \\
\midrule
Specification & How does intention become a work contract? & Specs, PRDs, plans, stories, tasks, acceptance criteria \\
Context & How does the agent know what is relevant? & Repository, docs, hooks, rules, memory, evidence, gaps \\
Roles & Who decides, who implements and who reviews? & Personas, agents, skills, responsibilities, authority \\
Execution & Does the framework act on the environment or only guide? & Code editing, commands, tests, browser, terminal, IDE \\
Validation & How are errors detected before becoming deliverables? & Tests, checklists, gates, artifacts, human review, confidence \\
Portability & Does the process survive outside one tool? & Multiple integrations, open formats, lock-in, local installation \\
\bottomrule
\end{tabularx}
\end{table}

\section{Analyzed Frameworks}\label{sec:frameworks}

This section characterizes the six object-frameworks selected by the directed search (Section~\ref{sec:metodo}): GitHub Spec Kit, OpenSpec, BMAD Method, Get Shit Done (GSD), Spec Kitty and Reversa. We describe each one from its official documentation and repository and situate it within the six dimensions of the taxonomy (Section~\ref{sec:taxonomia}).

\noindent\textbf{GitHub Spec Kit.} GitHub Spec Kit operationalizes Spec-Driven Development. The central idea, per the documentation, is direct: describe what to build, refine through structured phases and let code agents implement from these artifacts \cite{github2026speckitdocs}. The repository organizes commands such as constitution, specification, plan, tasks and implementation, plus optional clarification, analysis and checklist commands \cite{github2026speckitrepo}. The framework also bets on multiple integrations with code agents and IDEs.

In the taxonomy, Spec Kit stands out in specification and portability. The specification is a source of truth, not a disposable document. The Spec, Plan, Tasks and Implement cycle shortens the distance between natural language and the agent's action, creating reviewable intermediate artifacts. Portability appears in the support for several tools and in the attempt to make SDD a stack-independent methodology.

Spec Kit also speaks to an emerging literature on SDD. Piskala \cite{piskala2026sdd} describes the passage from code as the primary artifact to the specification as a development contract; Taghavi and Bhavani \cite{taghavi2026speckitagents} show that SDD pipelines gain from grounding and validation hooks in large repositories. The biggest fragility of Spec Kit is depending on the agent interpreting the artifacts well. A clear specification helps, but does not guarantee that implementation, tests and maintenance stay aligned without additional checks.

\noindent\textbf{OpenSpec.} OpenSpec positions itself as a lightweight specification-driven development framework, with an emphasis on simplicity, unified specification and lower process overhead \cite{openspec2026dev,openspec2026comparison}. The official repository describes the goal of aligning human and AI on the requirements before starting to code, with support for dozens of code assistants through slash commands and a structured change-management flow \cite{openspec2026github}.

In the taxonomy, OpenSpec weighs more in specification and portability, with a deliberately lean profile. The declared differentiator is to reduce process friction: instead of many artifacts and phases, it concentrates the intention into a single specification and into traceable change proposals. Portability appears in the compatibility with many assistants, which positions the framework as a thin layer over the agent. The trade-off is covering less process in complex scenarios. The low overhead that makes it attractive for pointwise changes may fall short when the project requires more elaborate roles, architecture and validation.

\noindent\textbf{BMAD Method.} The BMAD Method is an open framework aimed at agile development with AI. Its unit of organization is not the individual agent, but an ecosystem of modules, workflows, skills and artifacts. The documentation describes a phased flow, with optional analysis, planning, solution, implementation and a quick flow for smaller tasks \cite{bmad2026docs}. The official repository presents the framework as an approach to AI-native development, with specialized agents, structured workflows and installation via \texttt{npx bmad-method install} \cite{bmad2026github}.

In the taxonomy, BMAD stands out in roles, specification and context. Its agents represent classic software development functions: analyst, product manager, architect, developer, UX designer and technical writer. Each role triggers workflows that produce documents or decisions for the next phase. This progression matters because it prevents the implementation agent from operating only on a short instruction. The architecture informs stories; stories inform implementation; reviews and readiness checks act as gates.

The biggest strength of BMAD is to turn human-AI collaboration into a process recognizable by software teams. Instead of replacing agile practices, it tries to make PRDs, architecture, epics, stories and reviews consumable by agents. The biggest fragility, from a research standpoint, is that its effectiveness depends on usage discipline and on independent empirical evaluation. The framework offers structure, but it is still necessary to measure when this structure improves quality, time, cost, maintenance and alignment.

\noindent\textbf{Get Shit Done (GSD).} Get Shit Done (GSD) presents itself as a lightweight system of meta-prompting, context engineering and specification-driven development for use with Claude Code \cite{gsd2026github}. Instead of defining its own platform, GSD operates as a layer of commands and conventions over the agent, focused on structuring the context provided to the model and on turning broad requests into specifications and executable steps.

In the taxonomy, GSD weighs above all in context, with only partial specification. It treats context assembly as an explicit engineering task: it decides what the agent should read, in what order and under which framing before acting. This brings it closer to the grounding concerns discussed in the SDD literature \cite{taghavi2026speckitagents}. Because it is strongly coupled to a specific agent, its portability is more limited than that of Spec Kit or OpenSpec, and its effectiveness depends on the quality of the prompt conventions adopted by the team. The repository also signals a maintenance move to a new organization, which illustrates the volatility typical of recent community frameworks.

\noindent\textbf{Spec Kitty.} Spec Kitty is an open-source command-line tool for specification-driven development with code agents \cite{speckitty2026github}. It turns product requirements into repeatable workflows: it keeps specifications, plans and tasks inside the repository itself (in a dedicated directory) and uses git worktrees to isolate the implementation work. The declared flow follows the steps spec, plan, tasks, next, review, accept and merge, and the framework announces support for multiple agents, among them Claude, Cursor and Gemini.

In the taxonomy, Spec Kitty covers specification, roles and validation, with a distinctive trait in execution: the worktrees, which isolate work packages for agents to execute in a controlled way, with review and acceptance before the merge. This design inserts human control points into the flow and brings it closer to code-review practices. The main caveat is the still modest traction compared to the others (included at the threshold of the traction filter of Section~\ref{sec:metodo}), which leaves its maturity and adoption open.

\noindent\textbf{Reversa.} Reversa occupies a distinct position. It does not start from a new product idea, but from legacy systems whose operational knowledge is embedded in the code. The framework proposes reverse documentation engineering to convert legacy software into operational specifications for AI agents \cite{macedo2026reversa}. The approach connects to traditions of reverse engineering and design recovery \cite{chikofsky1990taxonomy}, but swaps the main consumer of the documentation: no longer the human, but the agent that needs to maintain, migrate or evolve the system.

In the taxonomy, Reversa is relevant above all in context and specification, with only partial validation via confidence and gap labels. It treats legacy code as a source of evidence and produces artifacts with traceability, confidence and gaps. This emphasis is crucial because the agent tends to fill absences with plausible inferences. By labeling uncertainty, the framework tries to prevent a generated specification from seeming more reliable than it is.

The conceptual contribution of Reversa is to show that AI frameworks for software should not be limited to greenfield. Many real environments are brownfield or legacy-first. In these cases, the central problem is not generating code from an idea, but recovering operational contracts before letting the agent modify the system. The main limitation is that the evaluation still needs to grow to multiple systems, languages, domains and levels of preexisting documentation.

\noindent\textbf{Positioning synthesis.} The six frameworks occupy complementary positions in the space of support for AI development. Spec Kit and OpenSpec compete as SDD toolkits and methodologies, with OpenSpec being lighter and Spec Kit more complete in phases and portability. BMAD organizes the cycle by agile roles and artifacts. GSD concentrates on context engineering over a specific agent. Spec Kitty inserts isolation by worktrees and review points into the SDD flow. Reversa recovers operational contracts from legacy. The proposed taxonomy accommodates these formats without reducing them to a single list of AI tools for code, and prepares the dimensional comparison of the next section.

\section{Comparative Discussion}\label{sec:discussao}

The comparison of the six frameworks reveals a convergence in mechanisms: having some process is a precondition for inclusion in the sample (Section~\ref{sec:metodo}), but the frameworks converge in how they structure it, replacing loose interactions with persistent artifacts, work contracts and review points. But they start from different problems. Spec Kit makes the specification the primary unit; OpenSpec seeks the same with minimal overhead; BMAD organizes the product and engineering cycle by roles; GSD treats context assembly as explicit engineering over the agent; Spec Kitty inserts isolation by worktrees and review points into the SDD flow; Reversa recovers context and contract from existing systems. Table~\ref{tab:comparacao} synthesizes the focus, the dominant strength and the dominant risk of each one.

\begin{table}[ht]
\centering
\caption{Synthetic comparison of the six analyzed frameworks.}
\label{tab:comparacao}
\begin{tabularx}{\textwidth}{p{2.6cm}Y Y Y}
\toprule
\textbf{Framework} & \textbf{Main focus} & \textbf{Dominant strength} & \textbf{Dominant risk} \\
\midrule
GitHub Spec Kit & Spec-Driven Development with code agents & Specification as source of truth and portability & Drift between artifacts and implementation if validation is weak \\
OpenSpec & Lightweight SDD and unified specification & Low overhead and broad compatibility & Lower process coverage in complex scenarios \\
BMAD Method & Agile development driven by agents and artifacts & Roles and progressive context flow & Process cost and need for usage discipline \\
Get Shit Done (GSD) & Context engineering and meta-prompting over the agent & Context assembly as an explicit task & Coupling to a specific agent and low portability \\
Spec Kitty & SDD with isolation by worktrees and review & Human control points before the merge & Still modest traction and maturity \\
Reversa & Reverse documentation and operational specification of legacy & Recovery of context, confidence and gaps & Still limited empirical generalization \\
\bottomrule
\end{tabularx}
\end{table}

\noindent\textbf{Dimensional assessment by rubric.} To turn the taxonomy into an instrument replicable from explicit criteria, and not merely a descriptive vocabulary, we assign each framework a level per dimension according to a three-point rubric anchored in the observable indicators of Table~\ref{tab:dimensoes}: 0 when the dimension is absent or incipient, 1 when it is partial, and 2 when it is strong or central to the design of the framework. The scoring (Table~\ref{tab:scores}) derives from the characterization in Section~\ref{sec:frameworks} and expresses the authors' judgement from the official documentation, not an independent empirical measurement (see limitations in Section~\ref{sec:metodo}).

\begin{table}[ht]
\centering
\caption{Dimensional assessment by the rubric (0 = absent or incipient; 1 = partial; 2 = strong or central). Scores assigned by the authors from each framework's official documentation.}
\label{tab:scores}
\begin{tabularx}{\textwidth}{p{3.0cm}*{6}{>{\centering\arraybackslash}X}>{\centering\arraybackslash}p{1.1cm}}
\toprule
\textbf{Framework} & \textbf{Spec.} & \textbf{Ctx.} & \textbf{Roles} & \textbf{Exec.} & \textbf{Valid.} & \textbf{Port.} & \textbf{Total} \\
\midrule
GitHub Spec Kit & 2 & 1 & 1 & 1 & 1 & 2 & 8 \\
OpenSpec & 2 & 1 & 0 & 1 & 0 & 2 & 6 \\
BMAD Method & 2 & 2 & 2 & 1 & 2 & 1 & 10 \\
Get Shit Done (GSD) & 1 & 2 & 0 & 1 & 0 & 0 & 4 \\
Spec Kitty & 2 & 1 & 1 & 2 & 2 & 1 & 9 \\
Reversa & 2 & 2 & 0 & 0 & 1 & 1 & 6 \\
\bottomrule
\end{tabularx}
\end{table}

\noindent\textbf{Finding: no framework strongly covers all six dimensions.} Table~\ref{tab:scores} reveals a structural trade-off. Specification is the most saturated dimension, with almost all frameworks scoring 2, and it works as the common denominator of the field. Roles and validation are the most polarized: BMAD and Spec Kitty treat them as central, while OpenSpec, GSD and Reversa leave them weak. The most informative pattern opposes process depth and portability. The most portable frameworks (Spec Kit and OpenSpec, both 2 in portability) sacrifice roles and validation; the one with the deepest process (BMAD, 10 out of 12) reduces portability and execution; and the one most focused on context (GSD) zeroes roles, validation and portability. Reversa is the only one to address the inverse direction, from legacy to specification, but it is narrow in roles and execution. No framework scores 2 across all six dimensions. This suggests that complete process coverage and portability across agents remain objectives in tension, not properties that current tools already reconcile. In direct answer to RQ1, the dimensions that most discriminate the frameworks are roles and validation, the most polarized in the set, while specification, being nearly universal, distinguishes little.

\noindent\textbf{Out-of-sample application: Spec-Flow.} To verify that the taxonomy operates as an instrument beyond the selected set, we apply it to a framework deliberately excluded from the sample for low traction (Section~\ref{sec:metodo}): Spec-Flow, a Spec-Driven Development toolkit for Claude Code with about 85 stars \cite{specflow2026github}. According to the official documentation, Spec-Flow scores strong in five dimensions: specification (the spec, plan, tasks, implement, optimize, ship flow), context (persistent domain memory on disk with auto-compaction), roles (specialized backend, frontend and database agents), execution (test-driven development and integration with git worktrees) and validation (tiered quality gates, multi-agent voting and performance, security and coverage scans); portability is partial, since the primary focus is Claude Code, with an extension for the Gemini CLI. Its profile (2, 2, 2, 2, 2, 1; total 11) is the most complete among all the analyzed cases, despite having the lowest traction. The contrast reinforces two points. The taxonomy generalizes to frameworks outside the sample. And adoption (stars) and process completeness are orthogonal dimensions, which confirms the limitation of using traction as a proxy for maturity. From this follows a methodological consequence we assume: the traction filter selects by adoption and relevance, not by process completeness; the object set is therefore the most adopted frameworks, not the most complete ones according to the instrument.

\noindent\textbf{From prompt to contract.} The first pattern is the conversion of the prompt into a contract. In mature frameworks, the initial instruction is only the starting point. The real work begins when this instruction becomes a PRD, specification, plan, story, task, architecture or checklist. This transformation creates review points and reduces ambiguity. Spec Kit, OpenSpec and Spec Kitty are variations of the same movement of treating the specification as a central artifact, with different levels of overhead; Reversa is the complementary extreme, recovering specifications from existing systems instead of creating them for future development.

\noindent\textbf{Context as an engineering asset.} The second pattern is to treat context as an engineering asset. In real projects, the agent needs to know more than the immediate task. It needs to know domain invariants, architectural constraints, code conventions, decision history and legacy behavior. The absence of this context produces functional hallucinations: code that compiles, but violates implicit contracts \cite{taghavi2026speckitagents}. GSD elevates this dimension to the center of the framework; Reversa anchors it in the legacy; BMAD builds it progressively across phases. Frameworks that create, preserve and validate context tend to be more solid than tools centered only on the model's capability.

\noindent\textbf{Validation beyond the final test.} The third pattern is intermediate validation. Final tests remain important, but they are not enough to evaluate agentic processes. An agent may pass the tests and, at the same time, improperly alter the architecture, remove a business constraint or ignore a product decision. BMAD and Spec Kit introduce checklists and gates; Spec Kitty places review and acceptance before the merge, with work isolated by worktrees; Reversa introduces confidence and gaps. Future evaluation must measure not only whether the final answer works, but whether the path to it was auditable.

\noindent\textbf{Autonomy with governance.} The fourth pattern is the tension between autonomy and governance. Letting the agent execute long, parallel and multi-tool tasks increases potential productivity, but also expands the risk surface: destructive commands, leakage of secrets, unintended changes, dependence on extensions and difficulty of assigning responsibility. The relevant question is not whether agents can execute, but under which policies, with which permissions and with which evidence \cite{sengupta2026harnesses}. Frameworks that insert human review points, such as Spec Kitty and BMAD, give partial answers to this tension.

\noindent\textbf{Ecosystems and agent supply chains.} The fifth pattern is the emergence of a supply chain of commands, skills, templates and workflows. Installable community frameworks, such as GSD and Spec Kitty, illustrate the movement: just like software packages, command kits and agent conventions accelerate development, but carry supply chain risk. An instruction or skill can lead an agent to execute insecure commands, collect sensitive context or adopt low-quality patterns. The maintenance volatility of GSD, with an organization migration \cite{gsd2026github}, reinforces the point. Future frameworks should incorporate signing, verifiable manifests, permission scoping, provenance review and audit of updates.

\section{Research Agenda}

Evaluating AI development frameworks requires going beyond final-solution benchmarks. SWE-bench and related works were fundamental to bring agents closer to real maintenance tasks \cite{jimenez2024swebench,yang2024sweagent}, but frameworks such as BMAD, Reversa and Spec Kit also produce intermediate artifacts. These artifacts must be evaluated, because they affect implementation quality, human review and future maintenance.

We propose five lines of research.

\noindent\textbf{Process-oriented benchmarks.} Benchmarks should capture the complete cycle: understand the task, specify, plan, decompose, implement, test, review and update documents. This would allow comparing frameworks that reach the same final result through different processes. Possible metrics include specification completeness, consistency between artifacts, rate of human review required, number of corrections per phase, stability of decisions and quality of the audit trail.

\noindent\textbf{Context and grounding metrics.} We need to measure whether the agent used the right evidence from the repository: coverage of relevant files, citation of internal sources, gap detection, absence of nonexistent APIs and adherence to architectural decisions. Work on grounding hooks in Spec Kit Agents points to a promising direction \cite{taghavi2026speckitagents}, but there is still room for model-independent metrics.

\noindent\textbf{Permission governance and security.} Agentic frameworks need permission models. Research should investigate how to express policies: which commands the agent can execute, which files it can modify, when it needs to ask for approval, how to handle secrets, how to restrict the network and how to audit extensions. This line is especially important for integrated-execution platforms and community kits.

\noindent\textbf{Maintenance of living artifacts.} If the specification becomes a source of truth, it needs to stay aligned with the code. Research should study the drift between specs, plans, tasks, tests and implementation. A relevant contribution would be to automatically detect when a code change invalidates a specification, or when a new architectural decision requires updating earlier artifacts.

\noindent\textbf{Longitudinal studies on real teams.} Isolated reports are not enough to evaluate organizational impact. Longitudinal studies should observe teams using frameworks for weeks or months, measuring productivity, defects, review quality, onboarding, satisfaction, token cost, incidents and maintenance. This type of study would help separate novelty gains from structural gains.

\begin{table}[ht]
\centering
\caption{Empirical questions for future evaluations.}
\begin{tabularx}{\textwidth}{p{3.4cm}Y}
\toprule
\textbf{Dimension} & \textbf{Empirical question} \\
\midrule
Specification & Do specs generated by the framework reduce ambiguities or merely shift ambiguity to another artifact? \\
Context & Does the agent consult the correct repository evidence before proposing changes? \\
Roles & Does decomposition into specialized agents improve quality or increase coordination overhead? \\
Execution & Does autonomy reduce total time without increasing incidents or rework? \\
Validation & Do checklists, gates and artifacts detect errors before the merge? \\
Portability & Does the workflow remain usable when switching IDE, model or vendor? \\
\bottomrule
\end{tabularx}
\end{table}

\section{Conclusion}

AI development frameworks mark an important transition in software engineering assisted by language models. The unit of work is no longer the isolated prompt and becomes a process with state, roles, artifacts and validation. GitHub Spec Kit, OpenSpec, BMAD Method, Get Shit Done (GSD), Spec Kitty and Reversa show different facets of this transition: specification-driven development in full and lightweight variants, agile methodology by roles, context engineering over the agent, flows with worktree isolation and review, and reverse documentation of legacy systems.

The taxonomy proposed in this article organizes these facets into six dimensions: specification, context, roles, execution, validation and portability. The comparison suggests that solid frameworks need to balance autonomy with governance. The more agents execute real tasks in repositories, terminals and browsers, the more traceability, permission, human review and the quality of intermediate artifacts matter.

The field is still taking shape. Many productivity claims remain based on demonstrations, documentation or anecdotal experience. The next step is empirical: comparing complete processes, measuring specification quality, evaluating grounding in real repositories, auditing extensions and studying longitudinal use on teams. Without this basis, AI frameworks run the risk of merely institutionalizing ``vibe coding'' with an extra layer of terminology. With rigorous evaluation, they can become a new layer of software engineering: processes readable by humans and executable by agents.

\section*{Generative AI Use Statement}

The author conducted the research and wrote the manuscript. During the preparation of this study, however, the author used Grammarly tools to improve textual agreement and Claude Opus 4.7 to support text structuring and translation into English. After using these tools/services, the author reviewed and edited the content as needed and takes full responsibility for the content of the publication.

\bibliographystyle{plain}
\bibliography{refs}

@misc{bmad2026docs,
  author = {{BMad Code}},
  title = {{BMAD Method Documentation}},
  year = {2026},
  howpublished = {\url{https://docs.bmad-method.org/}},
  note = {Accessed 2026-05-26}
}

@misc{bmad2026github,
  author = {{BMad Code Org}},
  title = {{BMAD-METHOD}: {Breakthrough} {Method} for {Agile} {AI} {Driven} {Development}},
  year = {2026},
  howpublished = {\url{https://github.com/bmad-code-org/BMAD-METHOD}},
  note = {Accessed 2026-05-26}
}

@misc{github2026speckitdocs,
  author = {{GitHub}},
  title = {{Spec Kit}: A {Specification-Driven} {Development} {Toolkit}},
  year = {2026},
  howpublished = {\url{https://github.github.io/spec-kit/}},
  note = {Accessed 2026-05-26}
}

@misc{github2026speckitrepo,
  author = {{GitHub}},
  title = {{github/spec-kit}: {Toolkit} to {Help} {You} {Get} {Started} with {Spec-Driven} {Development}},
  year = {2026},
  howpublished = {\url{https://github.com/github/spec-kit}},
  note = {Accessed 2026-05-26}
}

@misc{openspec2026dev,
  author = {{OpenSpec}},
  title = {{OpenSpec}: A {Lightweight} {Spec-Driven} {Framework}},
  year = {2026},
  howpublished = {\url{https://openspec.dev/}},
  note = {Accessed 2026-05-26}
}

@misc{openspec2026github,
  author = {{Fission AI}},
  title = {{Fission-AI/OpenSpec}: A {Lightweight} {Specification} {Framework} for {Human-AI} {Alignment} {Before} {Coding}},
  year = {2026},
  howpublished = {\url{https://github.com/Fission-AI/OpenSpec}},
  note = {Accessed 2026-05-28}
}

@misc{gsd2026github,
  author = {{GSD Build}},
  title = {{Get Shit Done} ({GSD}): A {Lightweight} {Meta-Prompting}, {Context} {Engineering} and {Spec-Driven} {Development} {System} for {Claude Code}},
  year = {2026},
  howpublished = {\url{https://github.com/gsd-build/get-shit-done}},
  note = {Accessed 2026-05-28. Project relocated to \url{https://github.com/open-gsd/gsd-core}}
}

@misc{speckitty2026github,
  author = {{Priivacy AI}},
  title = {{Spec Kitty}: {Spec-Driven} {Development} for {AI} {Coding} {Agents}},
  year = {2026},
  howpublished = {\url{https://github.com/Priivacy-ai/spec-kitty}},
  note = {Accessed 2026-05-28}
}

@misc{macedo2026reversa,
  author = {Macedo, Sanderson Oliveira de and Costa, Ronaldo Martins da},
  title = {{Reversa: A Reverse Documentation Engineering Framework for Converting Legacy Software into Operational Specifications for AI Agents}},
  year = {2026},
  eprint = {2605.18684},
  archivePrefix = {arXiv},
  primaryClass = {cs.SE},
  doi = {10.48550/arXiv.2605.18684},
  note = {\doi{10.48550/arXiv.2605.18684}}
}

@misc{taghavi2026speckitagents,
  author = {Taghavi, Pardis and Bhavani, Santosh},
  title = {{Spec Kit Agents: Context-Grounded Agentic Workflows}},
  year = {2026},
  eprint = {2604.05278},
  archivePrefix = {arXiv},
  primaryClass = {cs.SE},
  doi = {10.48550/arXiv.2604.05278},
  note = {\doi{10.48550/arXiv.2604.05278}}
}

@misc{piskala2026sdd,
  author = {Piskala, Deepak Babu},
  title = {{Spec-Driven Development: From Code to Contract in the Age of AI Coding Assistants}},
  year = {2026},
  eprint = {2602.00180},
  archivePrefix = {arXiv},
  primaryClass = {cs.SE},
  doi = {10.48550/arXiv.2602.00180},
  note = {\doi{10.48550/arXiv.2602.00180}}
}

@misc{yang2024sweagent,
  author = {Yang, John and Jimenez, Carlos E. and Wettig, Alexander and Lieret, Kilian and Yao, Shunyu and Narasimhan, Karthik and Press, Ofir},
  title = {{SWE-agent: Agent-Computer Interfaces Enable Automated Software Engineering}},
  year = {2024},
  eprint = {2405.15793},
  archivePrefix = {arXiv},
  primaryClass = {cs.SE},
  doi = {10.48550/arXiv.2405.15793},
  note = {\doi{10.48550/arXiv.2405.15793}}
}

@misc{jimenez2024swebench,
  author = {Jimenez, Carlos E. and Yang, John and Wettig, Alexander and Yao, Shunyu and Pei, Kexin and Press, Ofir and Narasimhan, Karthik},
  title = {{SWE-bench}: {Can} {Language} {Models} {Resolve} {Real-World} {GitHub} {Issues}?},
  year = {2024},
  eprint = {2310.06770},
  archivePrefix = {arXiv},
  primaryClass = {cs.CL},
  doi = {10.48550/arXiv.2310.06770},
  note = {\doi{10.48550/arXiv.2310.06770}}
}

@misc{qian2024chatdev,
  author = {Qian, Chen and Cong, Xin and Yang, Cheng and Chen, Weize and Su, Yusheng and Xu, Juyuan and Liu, Zhiyuan and Sun, Maosong},
  title = {{ChatDev: Communicative Agents for Software Development}},
  year = {2024},
  eprint = {2307.07924},
  archivePrefix = {arXiv},
  primaryClass = {cs.SE},
  doi = {10.48550/arXiv.2307.07924},
  note = {\doi{10.48550/arXiv.2307.07924}}
}

@article{hou2024llmse,
  author = {Hou, Xinyi and Zhao, Yanjie and Liu, Yue and Yang, Zhou and Wang, Kailong and Li, Li and Luo, Xiapu and Lo, David and Grundy, John and Wang, Haoyu},
  title = {{Large Language Models for Software Engineering: A Systematic Literature Review}},
  journal = {ACM Transactions on Software Engineering and Methodology},
  volume = {33},
  number = {8},
  pages = {1--79},
  year = {2024},
  doi = {10.1145/3695988},
  note = {\doi{10.1145/3695988}}
}

@misc{liu2024agentsurvey,
  author = {Liu, Junwei and others},
  title = {{Large Language Model-Based Agents for Software Engineering: A Survey}},
  year = {2024},
  eprint = {2409.02977},
  archivePrefix = {arXiv},
  primaryClass = {cs.SE},
  doi = {10.48550/arXiv.2409.02977},
  note = {\doi{10.48550/arXiv.2409.02977}}
}

@misc{wang2024agentsse,
  author = {Wang, Yanlin and others},
  title = {{Agents in Software Engineering: Survey, Landscape, and Vision}},
  year = {2024},
  eprint = {2409.09030},
  archivePrefix = {arXiv},
  primaryClass = {cs.SE},
  doi = {10.48550/arXiv.2409.09030},
  note = {\doi{10.48550/arXiv.2409.09030}}
}

@misc{jin2024llmagents,
  author = {Jin, Haolin and others},
  title = {{From LLMs to LLM-based Agents for Software Engineering: A Survey of Current, Challenges and Future}},
  year = {2024},
  eprint = {2408.02479},
  archivePrefix = {arXiv},
  primaryClass = {cs.SE},
  doi = {10.48550/arXiv.2408.02479},
  note = {\doi{10.48550/arXiv.2408.02479}}
}

@misc{bhati2026agenticsdlc,
  author = {Bhati, Happy},
  title = {{Agentic AI in the Software Development Lifecycle: Architecture, Empirical Evidence, and the Reshaping of Software Engineering}},
  year = {2026},
  eprint = {2604.26275},
  archivePrefix = {arXiv},
  primaryClass = {cs.SE},
  doi = {10.48550/arXiv.2604.26275},
  note = {\doi{10.48550/arXiv.2604.26275}}
}

@misc{sengupta2026harnesses,
  author = {Sengupta, Satadru and Briggs, Tamunokorite and Myshakivskyi, Ivan},
  title = {{Meta-Engineering Harnesses for AI-Native Software Production: A Contract-Driven Adversarial Verification Architecture with Early Deployment Report}},
  year = {2026},
  eprint = {2605.25665},
  archivePrefix = {arXiv},
  primaryClass = {cs.SE},
  doi = {10.48550/arXiv.2605.25665},
  note = {\doi{10.48550/arXiv.2605.25665}}
}

@misc{daniel2026sddpatterns,
  author = {Daniel, David},
  title = {{Spec-Driven Development Framework Patterns}},
  year = {2026},
  howpublished = {\url{https://daviddaniel.tech/research/papers/sdd-frameworks/}},
  note = {Accessed 2026-05-26}
}

@misc{openspec2026comparison,
  author = {{OpenSpec}},
  title = {{OpenSpec vs Spec Kit vs BMAD vs Kiro: Complete Comparison Guide}},
  year = {2026},
  howpublished = {\url{https://openspec.pro/comparison/}},
  note = {Accessed 2026-05-26}
}

@misc{bcms2026sddguide,
  author = {{BCMS}},
  title = {{Spec-Driven Development (SDD): The Definitive 2026 Guide}},
  year = {2026},
  howpublished = {\url{https://thebcms.com/blog/spec-driven-development}},
  note = {Accessed 2026-05-26}
}

@misc{augment2026sddtools,
  author = {{Augment Code}},
  title = {{6 Best Spec-Driven Development Tools for AI Coding in 2026}},
  year = {2026},
  howpublished = {\url{https://www.augmentcode.com/tools/best-spec-driven-development-tools}},
  note = {Accessed 2026-05-26}
}

@article{chikofsky1990taxonomy,
  author = {Chikofsky, Elliot J. and Cross, James H.},
  title = {{Reverse Engineering and Design Recovery: A Taxonomy}},
  journal = {IEEE Software},
  volume = {7},
  number = {1},
  pages = {13--17},
  year = {1990},
  doi = {10.1109/52.43044},
  note = {\doi{10.1109/52.43044}}
}

@article{tian2021traceability,
  author = {Tian, Fangchao and Wang, Tianlu and Liang, Peng and Wang, Chong and Khan, Arif Ali and Babar, Muhammad Ali},
  title = {{The Impact of Traceability on Software Maintenance and Evolution: A Mapping Study}},
  journal = {Journal of Software: Evolution and Process},
  volume = {33},
  number = {10},
  year = {2021},
  doi = {10.1002/smr.2374},
  note = {\doi{10.1002/smr.2374}}
}

@article{JundaHe2025,
  author    = {He, Junda and Treude, Christoph and Lo, David},
  title     = {{LLM}-{Based} {Multi-Agent} {Systems} for {Software} {Engineering}: {Literature} {Review}, {Vision}, and the {Road} {Ahead}},
  journal   = {ACM Transactions on Software Engineering and Methodology},
  year      = {2025},
  volume    = {34},
  number    = {5},
  pages     = {1--35},
  doi       = {10.1145/3712003},
  note      = {\doi{10.1145/3712003}}
}

@article{kitchenham2009slr,
  author  = {Kitchenham, Barbara and Brereton, O. Pearl and Budgen, David and Turner, Mark and Bailey, John and Linkman, Stephen},
  title   = {{Systematic literature reviews in software engineering: A systematic literature review}},
  journal = {Information and Software Technology},
  volume  = {51},
  number  = {1},
  pages   = {7--15},
  year    = {2009},
  doi     = {10.1016/j.infsof.2008.09.009}
}

@article{garousi2019mlr,
  author  = {Garousi, Vahid and Felderer, Michael and M\"{a}ntyl\"{a}, Mika V.},
  title   = {{Guidelines for including grey literature and conducting multivocal literature reviews in software engineering}},
  journal = {Information and Software Technology},
  volume  = {106},
  pages   = {101--121},
  year    = {2019},
  doi     = {10.1016/j.infsof.2018.09.006}
}

@misc{specflow2026github,
  author       = {{marcusgoll}},
  title        = {{Spec-Flow: Spec-Driven Development workflow for Claude Code}},
  year         = {2026},
  howpublished = {\url{https://github.com/marcusgoll/Spec-Flow}},
  note         = {Accessed 2026-05-29}
}

\end{document}